\documentclass[english,journal=cmatex,manuscript=article]{achemso}
\usepackage[latin9]{inputenc}
\usepackage{color}
\usepackage{units}
\usepackage{amstext}
\usepackage{graphicx}

\makeatletter

\title{\textcolor{black}{Crystal Structure of Silver Pentazolates Ag$\textnormal{\textbf{N}}{}_{\mathbf{5}}$
and Ag$\textnormal{\textbf{N}}{}_{\mathbf{6}}$}}
\author{\textcolor{black}{Ashley S. Williams, Kien Nguyen Cong, Joseph M.
Gonzalez, and Ivan I. Oleynik}}
\affiliation{\textcolor{black}{Department of Physics, University of South Florida,
Tampa, FL 33620}}
\email{oleynik@usf.edu}
\DeclareFontEncoding{LGR}{}{}
\DeclareRobustCommand{\greektext}{%
  \fontencoding{LGR}\selectfont\def\encodingdefault{LGR}}
\DeclareRobustCommand{\textgreek}[1]{\leavevmode{\greektext #1}}
\ProvideTextCommand{\~}{LGR}[1]{\char126#1}

\setkeys{acs}{articletitle=true}

\makeatother

\usepackage{babel}
\begin{document}
\begin{abstract}
\textcolor{black}{Silver pentazolate, a high energy density compound
containing cyclo-N$_{5}^{-}$ anion, has recently been synthesized
at ambient conditions. However, due to high sensitivity to irradiation,
its crystal structure has not been determined. In this work, silver-nitrogen
crystalline compounds at ambient conditions and high pressures, up
to 100 GPa, are predicted and characterized by performing first-principles
evolutionary crystal structure searching with variable stoichiometry.
It is found that newly discovered AgN$_{5}$ and AgN$_{6}$ are the
only thermodynamically stable silver-nitrogen compounds at pressures
between $42$ and $\unit[80]{GPa}$. In contrast to AgN$_{5}$, pentazolate
AgN$_{6}$ compound contains N$_{2}$ diatomic molecules in addition
to cyclo-N$_{5}^{-}$. These AgN$_{5}$ and AgN$_{6}$ crystals are
metastable at ambient conditions with positive formation enthalpies
of $\unit[54.95]{kJ/mol}$ and $\unit[46.24]{kJ/mol},$ respectively.
The underlying cause of the stability of cyclo-N$_{5}^{-}$ silver
pentazolates is the enhanced aromaticity enabled by the charge transfer
from silver atoms to nitrogen rings. To aid in experimental identification
of these materials, calculated Raman spectra are reported at ambient
pressure: the frequencies of N$_{5}^{-}$ vibrational modes of AgN$_{5}$
are in good agreement with those measured in experiment.}
\end{abstract}

\section{\textcolor{black}{Introduction}}

\textcolor{black}{Polynitrogen high energy density materials are in
the focus of intense experimental and theoretical investigations\cite{Steinhauser2008,Brinck2014}
due to their promise of delivering high explosive power upon decomposition
of single and double bonded nitrogen atoms to triply bonded N$_{2}$
\cite{Klapotke2007,Zarko2010}. Although pure nitrogen extended solids
such as cubic gauche (cg) and layered polymeric nitrogen have been
predicted \cite{Mailhiot1992} and subsequently synthesized at high
pressures \cite{Eremets2004a,Tomasino2014}, their recovery at ambient
conditions is challenging \cite{Eremets2008}.}

\textcolor{black}{Another possibility to achieve high energy density
is to realize compounds containing single and double bonded nitrogen
molecular species, such as cyclo-N$_{5}^{-}$ anion.\cite{Bazanov2016,Steele2016,Steele,Steele2017,Williams2017,Xu2017,Zhang2017a,Zhang2017b,Sun2018,Laniel2018}
The negative charge on pentazoles requires an electron donating component,
therefore, a family of alkali pentazolate crystals have been predicted
\cite{Williams2017,Steele,Steele2016,Steele2017,Steele2017a,Steele2019,Shen2015,Peng2015,Yi2020}
and then synthesized \cite{Bazanov2016,Steele2017,Xu2017,Zhang2017a,Zhang2017b,Sun2018,Laniel2018,Zhou2020,Bykov2021}
at high pressures. Success has recently been achieved in synthesizing
cyclo-N$_{5}^{-}$ pentazolates at ambient conditions using traditional
methods of synthetic chemistry \cite{Bazanov2016,Xu2017,Zhang2017a,Zhang2017b,Sun2018}.
However, the challenge of recovering simpler cyclo-N$_{5}^{-}$ pentazolate
compounds at ambient conditions still remains.}

\textcolor{black}{Most of metal pentazolates discovered so far are
the crystals with alkali cations Li$^{+}$\cite{Shen2015,Laniel2018,Zhou2020,Yi2020},
Na$^{+}$\cite{Steele2016,Bykov2021}, K$^{+}$\cite{Steele2017a},
and Rb$^{+}$\cite{Williams2017} with 1:5 metal/nitrogen stoichiometry.
They are just the one of several M$_{x}$N$_{y}$ materials featuring
isolated nitrogen atoms/ions, finite and infinite chains, 2D layers
and other polynitrogen complexes. Such rich chemistry is due to the
ease of alkali atoms to donate electrons to nitrogen species, resulting
in plethora of M$_{x}$N$_{y}$ chemical compositions.}

\textcolor{black}{Silver, being one of the coinage elements (Cu, Ag,
and Au), is ``noble'', i.e. almost unreactive, compared to reactive
alkali metals. This is because the outer valence electrons are more
tightly bound than those of alkali atoms due to poor screening of
the nuclear charge by the d- sub-shells compared to much stronger
screening by filled s/p shells of alkali atoms. Therefore, less variety
is expected among stoichiometries of Ag$_{x}$N$_{y}$ compounds;
it is highly probable that 1:5 stoichiometry would be the only one
that displays negative formation enthalpies at high pressures. In
fact, even the existence of AgN$_{5}$ was questioned more than a
century ago when it was purportedly synthesized by Lifschitz\cite{Lifschitz1915}
and then refuted by Curtius }\textit{\textcolor{black}{et al.}}\textcolor{black}{\cite{Curtius1915}
. Only two silver nitrogen crystalline compounds are known to exist
at ambient conditions: silver nitride Ag$_{3}$N\cite{Shanley1991}
and silver azide AgN$_{3}$\cite{Guo1999}, both of them being extremely
sensitive explosive materials\cite{Ennis1991,Kleine2003,Luchs1966}.}

\textcolor{black}{Sun et al. reported the synthesis of a solvent-free
pentazolate complex AgN$_{5}$ in a compound which does not contain
counter ions such as Cl$^{-}$, NH$_{4}{}^{+}$ and H$_{3}$O$^{{\color{red}+}}$
\cite{Sun2018}. The authors failed to obtain single crystal AgN$_{5}$
suitable to XRD characterization as this compound is highly unstable
upon irradiation or heating. Instead, they converted AgN$_{5}$ complex
to a 3D framework {[}Ag(NH$_{3}$)$_{2}${]}$^{+}${[}Ag$_{3}$(N$_{5}$)$_{4}${]}$^{-}$
in order to characterize its crystal structure. Therefore, the important
question is whether AgN$_{5}$ exists and if it does, what is its
crystal structure. The light and heat instability of AgN$_{5}$ indicates
a high degree of its metastability, similar to known silver azide
AgN$_{3}$ crystal which possesses a very high positive enthalpy of
formation $\unit[310.3]{kJ/mol}$ \cite{Gray1956}.}

\textcolor{black}{It is known that application of pressure imparts
a substantial thermomechanical energy of compression to precursor
materials\cite{Yoo2017}, which may promote unusual chemistry of resulting
compounds. Therefore, the investigation of AgN$_{5}$ as well as other
Ag-N materials with different stoichiometries (if they exist) at high
pressures is of great importance for elucidating bonding and structure
of silver-nitrogen compounds.}

\textcolor{black}{The goal of this research is to answer these outstanding
questions by performing crystal structure searching of new silver-nitrogen
crystalline compounds, including AgN$_{5}$, at ambient conditions
and high pressures up to 100 GPa using variable composition first-principles
evolutionary crystal structure prediction method. Once the new compounds
are uncovered their stability, bonding and structure as well as electronic
and vibrational properties are analyzed. Chemical pathways for synthesis
of discovered thermodynamically stable compounds are proposed and
Raman spectra of the resultant compounds are calculated to aid in
experimental detection of these materials at ambient pressure.}

\section{\textcolor{black}{Computational Details}}

\textcolor{black}{The silver-nitrogen compounds Ag$_{x}$N$_{y}$
of variable stoichiometry are searched using first-principles evolutionary
structure prediction method implemented in Universal Structure Predictor:
Evolutionary Xtallography code (USPEX) \cite{Oganov2006,Glass2006,Lyakhov2010}.
The variable composition searches are performed at several pressures
using unit cells containing varying amounts of silver and nitrogen
atoms between 8 to 16 atoms in the cell. In total, about 100 different
Ag$_{x}$N$_{y}$ stoichiometries are sampled. Once the variable composition
search at a given pressure is complete, fixed composition or molecular
searches using larger unit cell sizes up to 48 atoms are performed
at stoichiometries corresponding to the thermodynamically stable compounds
at the vertices of the convex hull and the metastable compounds up
to $\unit[20]{meV}$ above the convex hull.}

\textcolor{black}{The process of structure prediction begins by generating
structures based on all space groups and random lattice parameters
in the first generation. The obtained structures are then optimized
to a target pressure. Their normalized formation enthalpy is determined
as follows:}

\textcolor{black}{
\[
\triangle H_{Ag_{x}N_{y}}=(H_{Ag_{x}N_{y}}-xH_{Ag}-yH_{N})/(x+y),
\]
where $H_{Ag_{x}N_{y}}$, $H_{Ag}$ and $H_{N}$ are enthalpy of the
compounds and best structures of pure elements, respectively. Based
on the normalized formation enthalpy, the convex hull at each pressure
is constructed using the lowest formation enthalpy structures. For
each subsequent generation, a small subset of the lowest enthalpy
structures is kept while a new generation is built by adding more
random crystals and by applying variation operators to the best ``parent''
structures. The process is repeated until the best kept structures
do not change for ten consecutive generations.}

\textcolor{black}{Once combined variable and fixed stoichiometry searches
are completed at several pressures, all the resulting crystal structures
are combined in one single database, which is then used for high precision
geometry optimization and construction of the final convex hull at
every pressure of our interest, from 0 to 100 GPa. Such combined method
was shown to be successful in overcoming computational limitations
of the variable composition search. For example, during variable composition
search at 50 GPa only one thermodynamically stable stoichiometry,
AgN$_{5}$, was found. However, a second thermodynamically stable
stoichiometry AgN$_{6}$ was uncovered in subsequent fixed stoichiometry
molecular search using twice larger unit cell containing four silver
atoms and twenty-four nitrogen atoms.}

\textcolor{black}{The USPEX method and similar structure search methods
\cite{Pickard2009,Ma2008} have already been successful in prediction
of experimentally known crystal structures of pure elements. To speed
up the search, the known lowest enthalpy reference structures of pure
elements were taken: fcc Ag-}\textit{\textcolor{black}{Fm-3m}}\textcolor{black}{{}
for Ag across the entire pressure range, and \textgreek{a}-N$_{2}$
at $\unit[0]{GPa}$, \textgreek{e}-N$_{2}$ at $\unit[50]{GPa}$,
and cg-N above $\unit[50]{GPa}$ for reference structures of nitrogen.
If a compound lies on the convex hull it is considered thermodynamically
stable meaning it will not decompose into either the pure elements
or into a mixture of other Ag$_{x}$N$_{y}$ compounds \cite{Hermann2017}.
Compounds that lie above the convex hull must be dynamically stable,
i.e. no negative phonon frequencies, to be considered metastable.}

\textcolor{black}{Perdew-Burke-Ernzerhof (PBE) generalized gradient
approximation (GGA) to DFT\cite{Perdew1996} implemented in Vienna
ab initio simulation package VASP\cite{Kresse1996} is used to perform
the first-principles calculations. The projector augmented wave (PAW)
pseudopotentials are employed. During the USPEX search, each crystal
is optimized using a plane wave basis set with $\unit[500]{eV}$ energy
cutoff and k-point sampling density of $\unit[0.07]{\mathring{A}^{-1}}$.
At the completion of a search, the structures that are on the convex
hull as well as a subset of possible metastable structures with higher
formation energy, up to $\unit[100]{meV}$ above the convex hull,
are optimized using high accuracy setting to generate the accurate
convex hull. During the high accuracy calculations, the hard nitrogen
PAW pseudopotential is used to address a possibility of overlapping
cores of N atoms in triple bonded N$_{2}$. In addition, the energy
cutoff and k-point density are increased to $\unit[1000]{eV}$ and
$\unit[0.03]{\mathring{A}^{-1}}$ respectively. Frozen phonon approximation
is used to calculate the phonons at the gamma point as well as off-resonant
Raman frequencies.\cite{Lazzeri2003} The intensities are determined
by calculating the macroscopic dielectric polarizability tensors for
each normal mode eigenvector.\cite{Porezag1996} The atomic charges
and bond orders are calculated using density derived electrostatic
and chemical charges (DDEC) method implemented in VASP.\cite{Manz2010}}

\section{\textcolor{black}{Results and discussion}}

\textcolor{black}{At 0 GPa, no negative formation enthalpy Ag$_{x}$N$_{y}$
compounds exist, therefore, a concave hull consists of metastable
structures with positive formation enthalpies. This is in agreement
with the fact that at ambient conditions only two known silver-nitrogen
compounds, silver nitride (Ag$_{3}$N) and silver azide (AgN$_{3}$),
possess very high positive formation enthalpies.}

\begin{figure}[H]
\textcolor{black}{\includegraphics[scale=0.4]{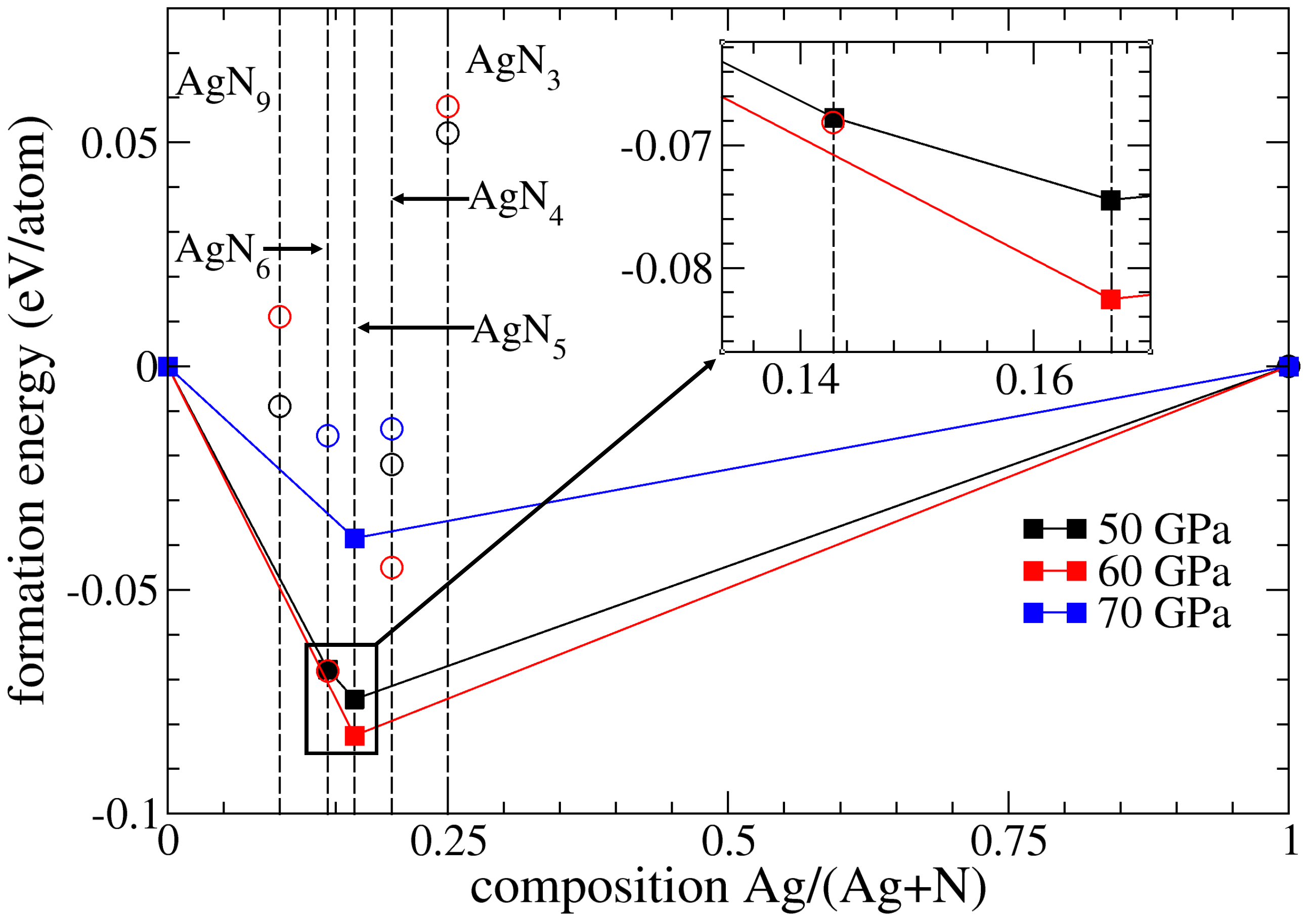}\caption{\label{fig:Convex-Hull}Ag-N convex hulls at 50, 60, and 70 GPa. Filled
squares indicate thermodynamically stable structures; open circles
-- metastable structures.}
}
\end{figure}

\textcolor{black}{At $\unit[50]{GPa}$, the stoichiometries with negative
formation enthalpy are thermodynamically stable AgN$_{5}$, and AgN$_{6}$,
and metastable AgN$_{9}$, and AgN$_{4}$, see the convex hull in
Fig. \ref{fig:Convex-Hull}. At $\unit[60]{GPa}$ and $\unit[70]{GPa}$
only AgN$_{5}$-}\textit{\textcolor{black}{P2$_{1}$/c}}\textcolor{black}{{}
is on the convex hull while AgN$_{6}$-}\textit{\textcolor{black}{P2$_{1}$2$_{1}$2}}\textcolor{black}{{}
becomes metastable with $\unit[1.93]{meV/atom}$ and $\unit[17.4]{meV/atom}$
above the convex hull at each pressure respectively.}

\textcolor{black}{Both AgN$_{5}$-}\textit{\textcolor{black}{P2$_{1}$/c}}\textcolor{black}{{}
and AgN$_{6}$-}\textit{\textcolor{black}{P2$_{1}$2$_{1}$2}}\textcolor{black}{{}
follow the pattern of NaN$_{5}$ which contains a single phase of
XN$_{5}$ over the entire pressure range\cite{Steele2016} whereas
KN$_{5}$, RbN$_{5},$ and CsN$_{5}$ consist of a low pressure and
a high pressure phases.\cite{Steele,Steele2016,Steele2017a,Williams2017}
The formation enthalpy of AgN$_{5}$-}\textit{\textcolor{black}{P2$_{1}$/c
}}\textcolor{black}{becomes negative for the first time at $\unit[41.5]{GPa}$.
At $\unit[57.5]{GPa}$, AgN$_{5}$-}\textit{\textcolor{black}{P2$_{1}$/c}}\textcolor{black}{{}
reaches its lowest formation enthalpy of $\unit[-0.095]{eV/atom}$.
Finally, at $\unit[79.4]{GPa}$, the formation enthalpy of AgN$_{5}$-}\textit{\textcolor{black}{P2$_{1}$/c
}}\textcolor{black}{becomes positive again. Similarly, the formation
enthalpy of AgN$_{6}$-}\textit{\textcolor{black}{P2$_{1}$2$_{1}$2}}\textcolor{black}{{}
becomes negative at $\unit[41.7]{GPa}$. At $\unit[57.5]{GPa}$, the
formation enthalpy AgN$_{6}$-}\textit{\textcolor{black}{P2$_{1}$2$_{1}$2}}\textcolor{black}{{}
reaches the lowest value of $\unit[-0.086]{eV/atom}$. The formation
enthalpy of AgN$_{6}$-}\textit{\textcolor{black}{P2$_{1}$2$_{1}$2}}\textcolor{black}{{}
becomes positive again at pressures exceeding $\unit[68]{GPa}$.}

\textcolor{black}{}

\textcolor{black}{}
\begin{figure}[H]
\textcolor{black}{\includegraphics[scale=0.4]{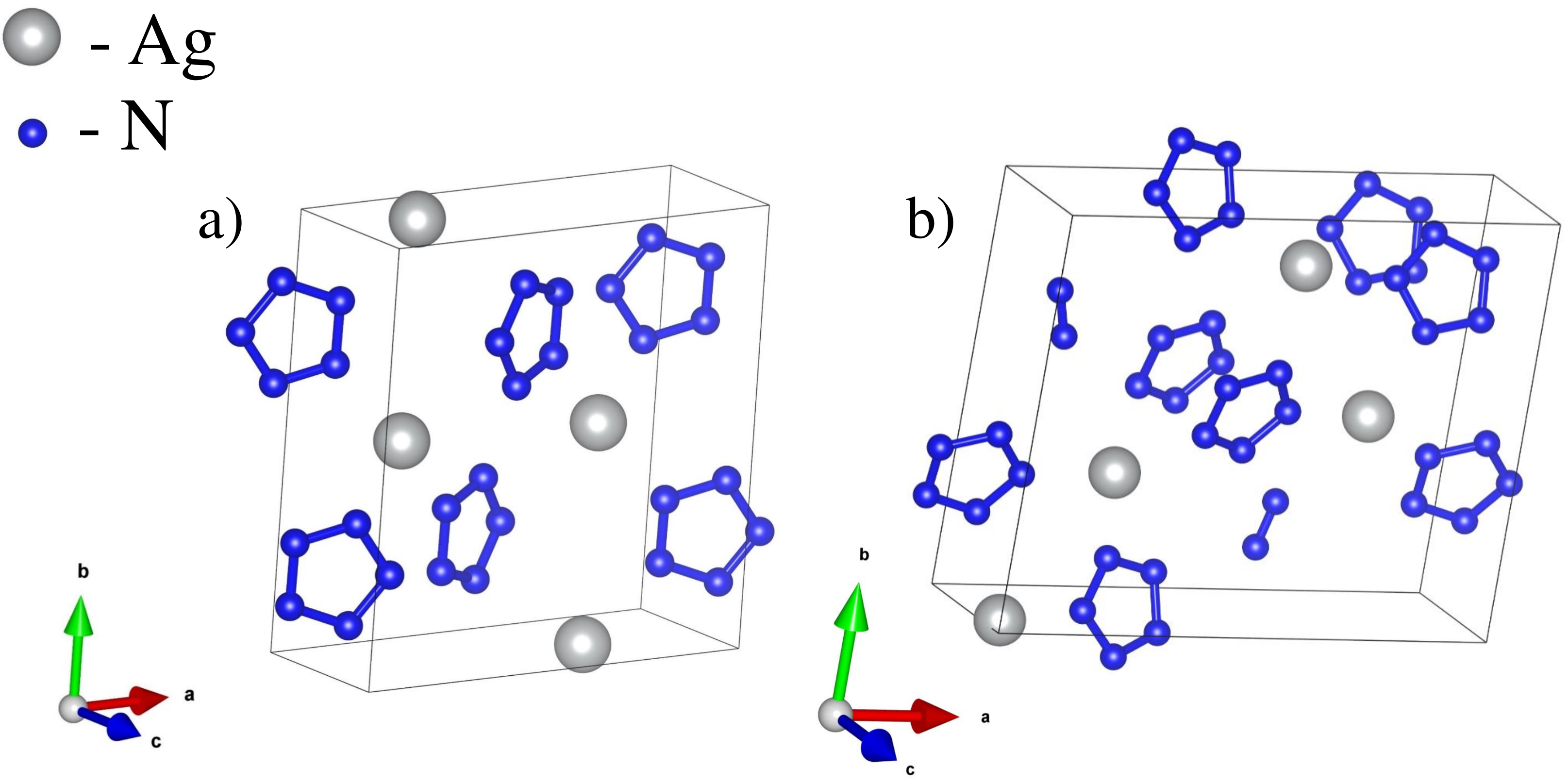}\caption{\label{fig:AgN5-AgN6-structures}Crystal structure of a) AgN$_{5}$-\textit{P2$_{1}$/c
}and b) AgN$_{6}$-\textit{P2$_{1}$2$_{1}$2 }\textit{\emph{pentazolates}}\textit{.
}The Ag atoms are depicted in gray and the N atoms are depicted in
blue.}
}
\end{figure}

\textcolor{black}{The crystal structures of the two new compounds
are depicted in Fig. \ref{fig:AgN5-AgN6-structures}. There are 24
atoms in the AgN$_{5}$-}\textit{\textcolor{black}{P2$_{1}$/c }}\textcolor{black}{unit
cell and 28 atoms in the AgN$_{6}$-}\textit{\textcolor{black}{P2$_{1}$2$_{1}$2}}\textcolor{black}{{}
unit cell. In AgN$_{5}$-}\textit{\textcolor{black}{P2$_{1}$/c,}}\textcolor{black}{{}
nitrogen atoms form 5 membered cyclo-N$_{5}^{-}$ pentazolate rings
with the silver atoms not covalently bonded to the nitrogen rings
or other silver atoms. The packing efficiency of AgN$_{5}$-}\textit{\textcolor{black}{P2$_{1}$/c}}\textcolor{black}{{}
is $\unit[43.8\%]{}$. The distance between the molecular centers
of nearest neighboring cyclo-N$_{5}^{-}$ pentazolate rings in AgN$_{5}$-}\textit{\textcolor{black}{P2$_{1}$/c}}\textcolor{black}{{}
is approximately $\unit[5]{\mathring{A}}$. In AgN$_{6}$-}\textit{\textcolor{black}{P2$_{1}$2$_{1}$2}}\textcolor{black}{,
nitrogen atoms form cyclo-N$_{5}^{-}$ with extra, isolated N$_{2}$
dimers and standalone silver atoms. The packing efficiency of AgN$_{6}$-}\textit{\textcolor{black}{P2$_{1}$2$_{1}$2}}\textcolor{black}{{}
is $\unit[24.6\%]{}$. The molecular centers of nearest neighboring
cyclo-N$_{5}^{-}$ pentazolate rings in AgN$_{6}$-}\textit{\textcolor{black}{P2$_{1}$2$_{1}$2}}\textcolor{black}{{}
are separated by approximately $\unit[5.2]{\mathring{A}}$. The cyclo-N$_{5}^{-}$
pentazolate rings of both AgN$_{5}$-}\textit{\textcolor{black}{P2$_{1}$/c}}\textcolor{black}{{}
and AgN$_{6}$-}\textit{\textcolor{black}{P2$_{1}$2$_{1}$2}}\textcolor{black}{{}
are isolated with bonding only occurring between the nitrogen atoms
within each individual ring. The electronic band structure of both
predicted silver pentazolate crystals display insulator character
at all pressures studied, see Figs. S1 and S2 in the supplementary
information.}

\textcolor{black}{}
\begin{figure}[H]
\textcolor{black}{\includegraphics[scale=0.5]{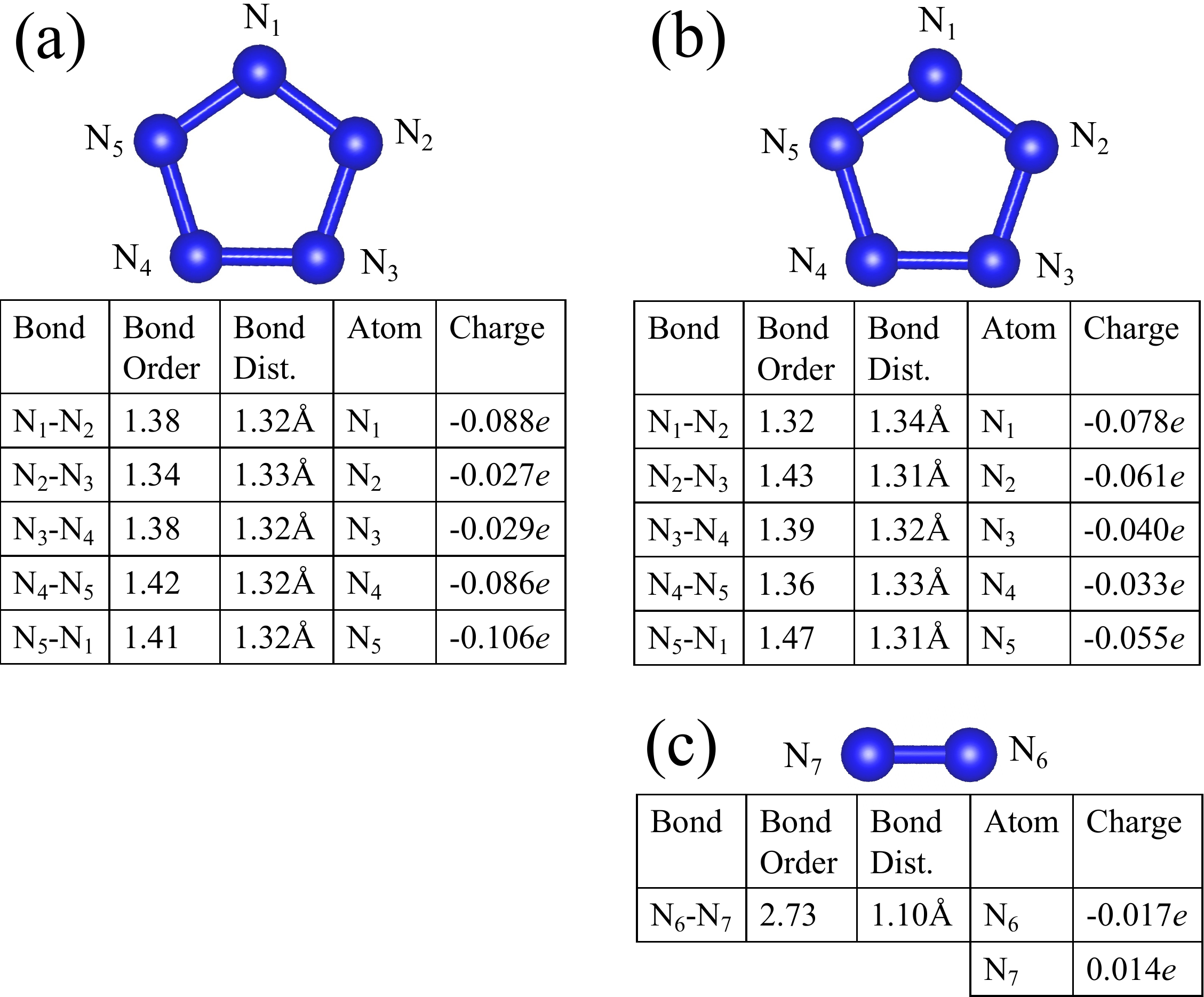}}

\textcolor{black}{\caption{\label{fig:Charges-and-Bond-Order} a) The charges and bond order
in the cyclo-N$_{5}^{-}$ pentazolate in the AgN$_{5}$-\textit{P2$_{1}$/c}
crystal. The N-N bond lengths in cyclo-N$_{5}^{-}$ pentazolate between
each silver atom is between $1.33\text{ Å}$ and $\unit[1.33\:]{\mathring{A}}$.
b) The charges and bond order in the cyclo-N$_{5}^{-}$ pentazolate
in the AgN$_{6}$-\textit{P2$_{1}$2$_{1}$2} crystal. The N-N bond
lengths in the cyclo-N$_{5}^{-}$ pentazolate is between $\unit[1.31]{}$
and $\unit[1.34]{\mathring{A}}$. c) The charges and bond order in
the N$_{2}$ dimer in the AgN$_{6}$-\textit{P2$_{1}$2$_{1}$2} crystal:
the N-N bond length is $\unit[1.10]{\mathring{A}}$.}
}
\end{figure}

\textcolor{black}{To understand the bonding of these novel compounds,
the charges and bond orders were calculated, see Fig. \ref{fig:Charges-and-Bond-Order}.
The N-N bond lengths in the pentazolate cyclo-N$_{5}^{-}$ anions
in AgN$_{5}$-}\textit{\textcolor{black}{P2$_{1}$/c}}\textcolor{black}{{}
are between $\unit[1.32]{\mathring{A}}$ and $\unit[1.33]{\mathring{A}}$.
The charges, beginning from the atom at the top point of the pentazolate
ring shown in Fig. \ref{fig:Charges-and-Bond-Order}(a) and moving
clockwise, are: -0.088}\textit{\textcolor{black}{e}}\textcolor{black}{,
-0.027}\textit{\textcolor{black}{e}}\textcolor{black}{, -0.029}\textit{\textcolor{black}{e}}\textcolor{black}{,
-0.086}\textit{\textcolor{black}{e}}\textcolor{black}{, and -0.106}\textit{\textcolor{black}{e}}\textcolor{black}{.
The silver atom carries charge +0.33}\textit{\textcolor{black}{e,
}}\textcolor{black}{the corresponding negative charge is on N$_{5}^{-}$
anion}\textit{\textcolor{black}{.}}\textcolor{black}{{} The aromatic
nature of the cyclo-N$_{5}^{-}$ can be deduced from the N-N bond
orders which range from $\unit[1.34]{}$ to $\unit[1.42]{}$, i.e.
between those corresponding to single and and double bonds. The average
charge and average N-N bond order in the cyclo-N$_{5}^{-}$ in AgN$_{5}$-}\textit{\textcolor{black}{P2$_{1}$/c
}}\textcolor{black}{are similar to their counterparts in the cyclo-N$_{5}^{-}$
in XN$_{5}$ \{X = Na, K, Rb, Cs\} compounds.\cite{Steele2016,Steele,Steele2017a,Williams2017}}

\textcolor{black}{The N-N bond lengths in cyclo-N$_{5}^{-}$ in AgN$_{6}$-}\textit{\textcolor{black}{P2$_{1}$2$_{1}$2
crystal}}\textcolor{black}{{} are between $\unit[1.31]{}$ and $\unit[1.34]{\mathring{A}}$
which are slightly smaller than those of its pure pentazolate counterpart,
AgN$_{5}$-}\textit{\textcolor{black}{P2$_{1}$/c. }}\textcolor{black}{As
seen in Fig. \ref{fig:Charges-and-Bond-Order}(b), the charges on
the nitrogen atoms in the AgN$_{6}$-}\textit{\textcolor{black}{P2$_{1}$2$_{1}$2}}\textcolor{black}{{}
pentazolate are also smaller on average than those in AgN$_{5}$-}\textit{\textcolor{black}{P2$_{1}$/c}}\textcolor{black}{{}
crystal, ranging from $\unit[-0.033e]{}$ to $\unit[-0.078e]{}$.
The N$_{2}$ dimer of AgN$_{6}$-}\textit{\textcolor{black}{P2$_{1}$2$_{1}$2}}\textcolor{black}{{}
is almost electroneutral, as the atomic charges on N atoms $\unit[-0.017e]{}$
and $\unit[0.014e]{}$ compensate each other, see Fig. \ref{fig:Charges-and-Bond-Order}c.
The silver atoms in AgN$_{6}$-}\textit{\textcolor{black}{P2$_{1}$2$_{1}$2}}\textcolor{black}{{}
carry a charge of $\unit[0.27e]{}$. Though the charges are smaller
in AgN$_{6}$-}\textit{\textcolor{black}{P2$_{1}$2$_{1}$2}}\textcolor{black}{{}
crystal than in AgN$_{5}$-}\textit{\textcolor{black}{P2$_{1}$/c}}\textcolor{black}{,
the N-N bond orders are comparable, ranging from $\unit[1.32]{}$
to $\unit[1.47]{}$ in case of AgN$_{6}$-}\textit{\textcolor{black}{P2$_{1}$2$_{1}$2}}\textcolor{black}{.
The N-N bond of the N$_{2}$ dimer in AgN$_{6}$-}\textit{\textcolor{black}{P2$_{1}$2$_{1}$2
}}\textcolor{black}{is close to a triple bond in isolated diatomic
N$_{2}$ molecule with N-N bond length $\unit[1.1]{\mathring{A}}$
and bond order $\unit[2.73]{}$.}

\textcolor{black}{}
\begin{figure}[H]
\textcolor{black}{\includegraphics[scale=0.4]{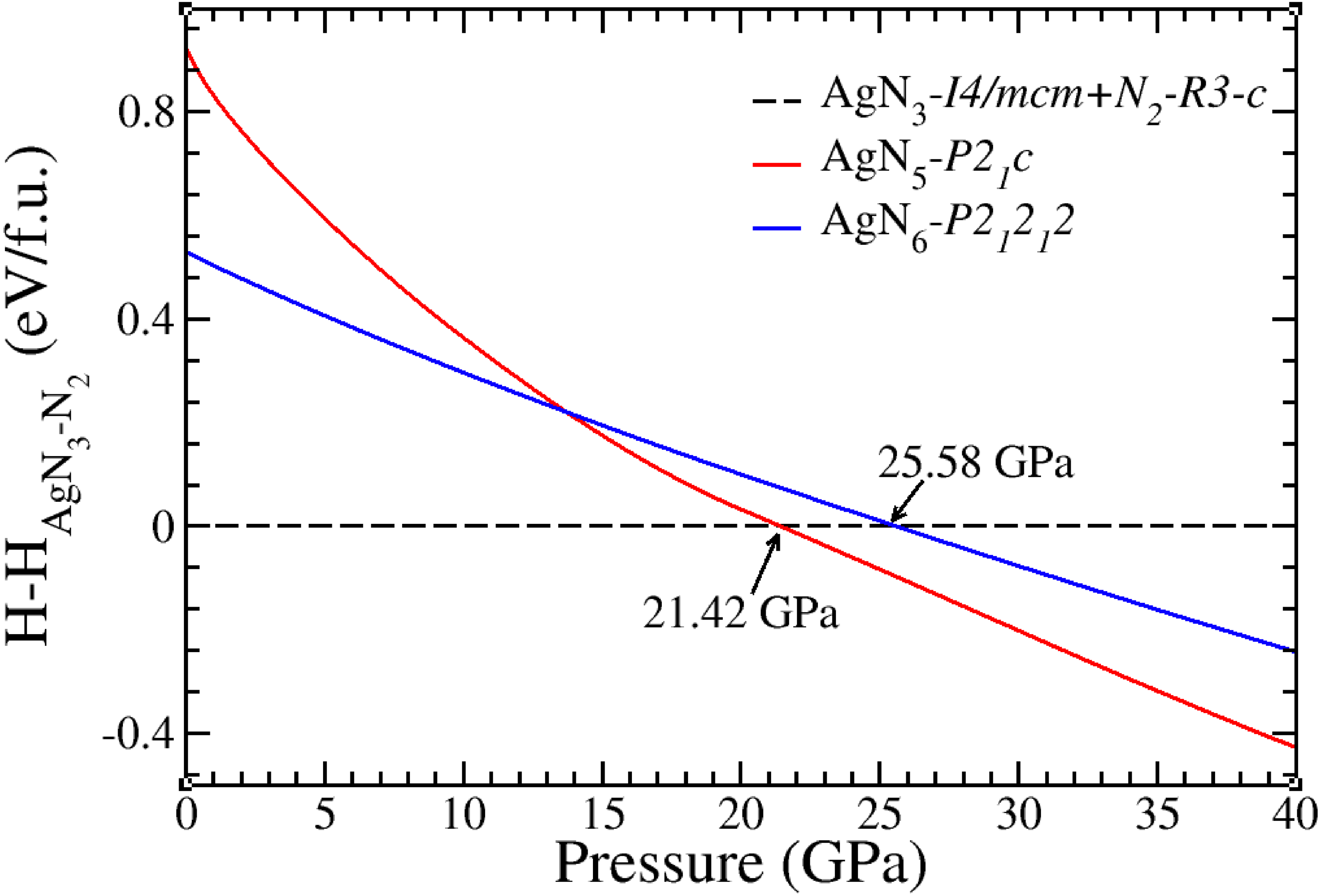}\caption{\label{fig:Compression}Enthalpy difference as a function of pressure
between AgN$_{5}$-\textit{P2$_{1}$/c }(red) and AgN$_{6}$-\textit{P2$_{1}$2$_{1}$2}
(blue) and a mixture of possible precursors, silver azide AgN$_{3}$
and nitrogen N$_{2}$.}
}
\end{figure}

\textcolor{black}{In principle, the new compounds can be synthesized
via compression of silver azide and nitrogen in a diamond anvil cell
(DAC). Several attempts have been made to produce pentazolate salts
via traditional DAC experiments whereby a combination of a metal azide
and molecular nitrogen is cold compressed to high pressures and then
laser heated. CsN$_{5}$ was synthesized, having first been guided
by first-principles crystal structure prediction, by Steele }\textit{\textcolor{black}{et
al.}}\textcolor{black}{\cite{Steele} via compression of CsN$_{5}$
+ N$_{2}$ in DAC to $\unit[60]{GPa}$. Zhou }\textit{\textcolor{black}{et
al.}}\textcolor{black}{\cite{Zhou2020} compressed lithium azide and
molecular nitrogen in DAC to pressure of only $\unit[41.1]{GPa}$
and report the formation of a new phase of LiN$_{5}$-}\textit{\textcolor{black}{P2$_{1}$/m.
}}\textcolor{black}{Zhou }\textit{\textcolor{black}{et al.'s}}\textcolor{black}{{}
results are consistent with predictions of the transition pressure
of $\unit[40]{GPa}$ by Yi }\textit{\textcolor{black}{et al.}}\textcolor{black}{\cite{Yi2020}
and Peng }\textit{\textcolor{black}{et al.}}\textcolor{black}{\cite{Peng2015}.
Recently, the formation of two new pentazolate salts, NaN$_{5}$ and
NaN$_{5}$·N$_{2}$, were observed via DAC experiments of sodium azide
and molecular nitrogen compressed to \textasciitilde$\unit[50]{GPa}$ .\cite{Bykov2021}
Notably, the NaN$_{5}$·N$_{2}$ crystal synthesized by Bykov }\textit{\textcolor{black}{et
al.}}\textcolor{black}{{} contains both cyclo-N$_{5}^{-}$ pentazolate
rings and N$_{2}$ dimers similar to what is predicted here for AgN$_{6}$-}\textit{\textcolor{black}{P2$_{1}$2$_{1}$2}}\textcolor{black}{.}

\textcolor{black}{Silver azide, in spite of its volatility, is well
studied with a known phase transition from low pressure AgN$_{3}$-}\textit{\textcolor{black}{Ibam}}\textcolor{black}{{}
to high pressure AgN$_{3}$-}\textit{\textcolor{black}{I4/mcm }}\textcolor{black}{at
$\unit[2.7]{GPa}$.\cite{Hou2011,Li2016} The theoretical study presented
by W. Zhu }\textit{\textcolor{black}{et al}}\textcolor{black}{. predicts
this transition to occur at nearly $\unit[7]{GPa}$\cite{Zhu2007a}
however, our results are closer to what was found in experiments with
AgN$_{3}$-}\textit{\textcolor{black}{I4/mcm}}\textcolor{black}{{} becoming
energetically favorable near $\unit[1.5]{GPa}$.}

\textcolor{black}{The enthalpy difference between AgN$_{5}$-}\textit{\textcolor{black}{P2$_{1}$/c
}}\textcolor{black}{and AgN$_{6}$-}\textit{\textcolor{black}{P2$_{1}$2$_{1}$2}}\textcolor{black}{{}
and the possible precursors, AgN$_{3}$-}\textit{\textcolor{black}{I4/mcm}}\textcolor{black}{{}
and N$_{2}$-}\textit{\textcolor{black}{R3-c}}\textcolor{black}{,
was calculated over increasing pressure as a zero-temperature estimate
of potential transformation of the precursors to the pentazolate product,
see Fig. \ref{fig:Compression}. AgN$_{3}$-}\textit{\textcolor{black}{I4/mcm}}\textcolor{black}{{}
is taken as the precursor due to the phase transition from AgN$_{3}$-}\textit{\textcolor{black}{Ibam}}\textcolor{black}{{}
to AgN$_{3}$-}\textit{\textcolor{black}{I4/mcm}}\textcolor{black}{{}
that occurs at such a low pressure. The results show that AgN$_{5}$-}\textit{\textcolor{black}{P2$_{1}$/c}}\textcolor{black}{{}
and AgN$_{6}$-}\textit{\textcolor{black}{P2$_{1}$2$_{1}$2}}\textcolor{black}{{}
are thermodynamically favorable to the mixture of the precursors at
$0\:\mathrm{K}$ and pressures exceeding $\unit[21]{GPa}$ and $\unit[26]{GPa}$
respectively.}

\textcolor{black}{It is important to note that $0\:\mathrm{K}$ data
are the lowest pressure estimates. The transition pressures at high
temperatures are determined by a complex interplay of thermodynamics
and kinetics of solid-state reactions.\cite{Erba2011} On the one
hand, temperature helps to activate kinetics, thus potentially lowering
transition pressure. \cite{Plasienka2015} On the other hand, temperature
might thermodynamically stabilize the molecular precursors compared
to solid reactants due to entropic -TS contribution to the Gibbs free
energy, thus increasing transition pressure upon increase of temperature.\cite{Erba2011,Alkhaldi2019}
Due to extreme complexity of the phenomena involved a meaningful predictions
of finite temperature effects seem to be impractical to make within
the scope of the current work.}

\textcolor{black}{}
\begin{figure}[H]
\textcolor{black}{\includegraphics[scale=0.4]{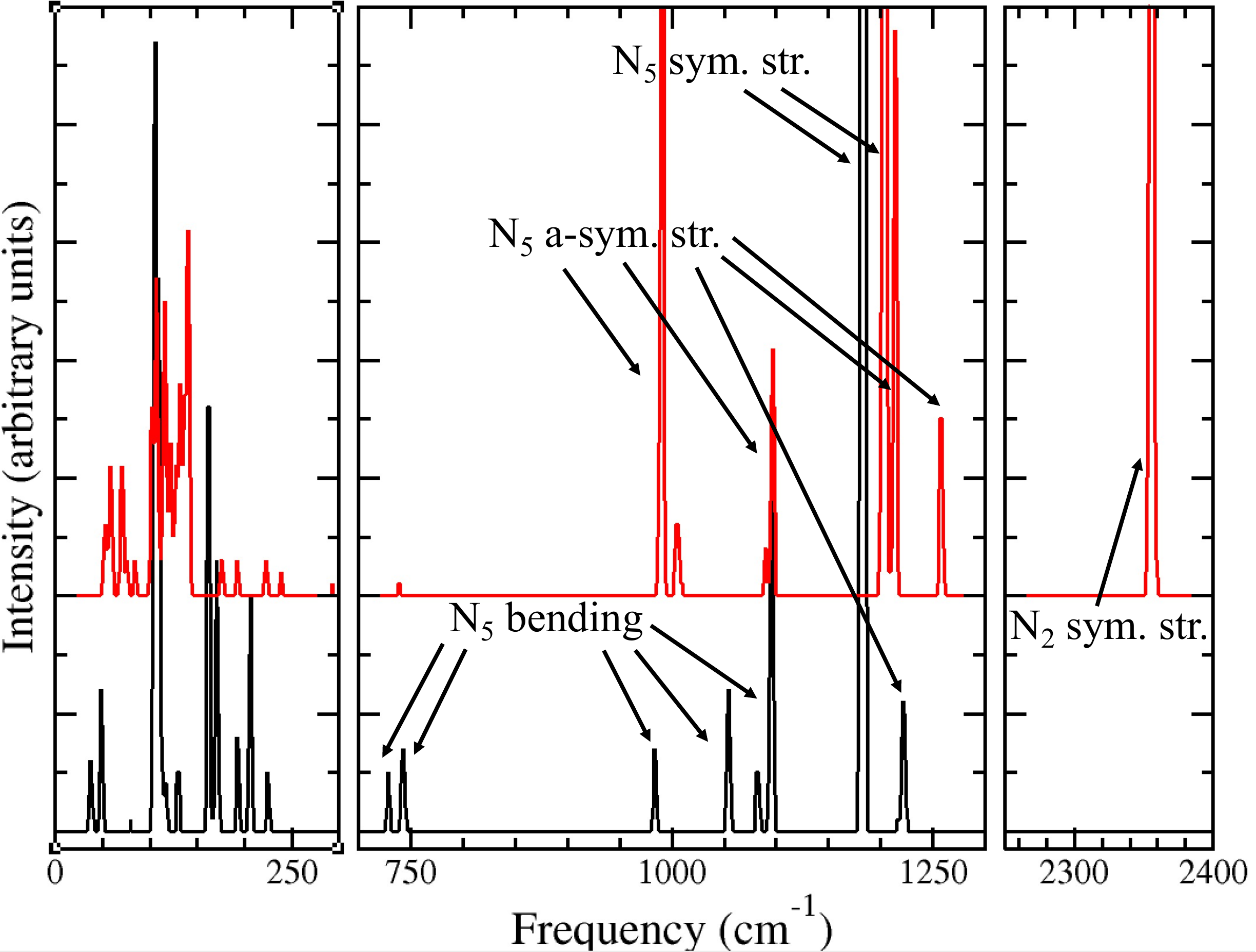}}

\textcolor{black}{\caption{\label{fig:Raman-Combined}Calculated Raman spectra of AgN$_{5}$-\textit{P2$_{1}$/c},
black curve, and AgN$_{6}$-\textit{P2$_{1}$2$_{1}$2}, red curve,
at $\unit[0]{GPa}$ zoomed to show smaller peaks. The tallest peak
of AgN$_{5}$-\textit{P2$_{1}$/c} at $\unit[1183.4]{cm^{-1}}$ is
nearly $\unit[7]{}$ times larger than the next tallest peak at \textasciitilde$\unit[106]{cm^{-1}}$ .
The tallest peak of AgN$_{6}$-\textit{P2$_{1}$2$_{1}$2} at $\unit[1200.2]{cm^{-1}}$
corresponding to the symmetric stretching of the N$_{5}^{-}$ molecule
is $\unit[1.42]{}$ times larger than the next tallest peak at $\unit[2355.9]{cm^{-1}}$
corresponding to the symmetric stretching of the N$_{2}$ dimer. The
specific bending and stretching modes are labeled in the high frequency
range.}
}
\end{figure}

\textcolor{black}{A material must be at least dynamically stable at
ambient conditions to be considered as a viable candidate for experimental
synthesis. The dynamical stability of AgN$_{5}$-}\textit{\textcolor{black}{P2$_{1}$/c
}}\textcolor{black}{at $\unit[0]{GPa}$ is confirmed via the absence
of negative phonon frequencies in phonon band structure, see Fig.
S3 in the supplementary information. To aid in the experimental identification
of AgN$_{5}$-}\textit{\textcolor{black}{P2$_{1}$/c}}\textcolor{black}{,
its Raman spectrum is calculated at $\unit[0]{GPa}$, see Fig. \ref{fig:Raman-Combined}.
The lattice and N$_{5}^{-}$ librational modes are in the low frequency
range of the Raman spectra, i.e. from $\unit[37.6]{}$ to $\unit[225.4]{cm^{-1}}$.
There are six bending modes in the Raman spectra: two twisting modes
at $\unit[728.4]{}$ and $\unit[742.6]{cm^{-1}}$ and four scissoring
modes at $\unit[984.2]{}$, $\unit[1054.8]{}$, $\unit[1083]{}$,
and $\unit[1101.3]{cm^{-1}}$. The most intense peak is the symmetric
stretching mode at $\unit[1183.4]{cm^{-1}}$ while the last mode at
$\unit[1222]{cm^{-1}}$ is an anti-symmetric stretching mode. These
results appear to agree well with experimental Raman spectra presented
by Sun }\textit{\textcolor{black}{et al.}}\textcolor{black}{\cite{Sun2018}.
Specifically, both structures show symmetric stretching modes, with
the mode reported in this study at $\unit[1183.4]{cm^{-1}}$ and Sun
}\textit{\textcolor{black}{et al.}}\textcolor{black}{{} reporting theirs
at $\unit[1187]{cm^{-1}}$.\cite{Sun2018} There are also two bending
modes reported in each study that are close in frequency, those at
$\unit[1054.8]{}$ and $\unit[1101.3]{cm^{-1}}$ in this study and
at $\unit[1020]{}$ and $\unit[1122]{cm^{-1}}$in Sun }\textit{\textcolor{black}{et
al.}}\textcolor{black}{'s work.\cite{Sun2018}}

\textcolor{black}{Negative phonon frequencies at ambient conditions
are also not present in case of AgN$_{6}$-}\textit{\textcolor{black}{P2$_{1}$2$_{1}$2}}\textcolor{black}{,
see Fig. S4 in the supplementary information. The Raman spectra of
this crystal are also calculated at $\unit[0]{GPa}$ to aid in its
experimental identification, see Fig. \ref{fig:Raman-Combined}. The
lattice and N$_{5}^{-}$ librational modes are found in the low frequency
range of $\unit[55.4]{}$ to $\unit[293.2]{cm^{-1}}$. Unlike AgN$_{5}$-}\textit{\textcolor{black}{P2$_{1}$/c}}\textcolor{black}{,
there are no N$_{5}^{-}$ bending modes present in AgN$_{6}$-}\textit{\textcolor{black}{P2$_{1}$2$_{1}$2}}\textcolor{black}{.
The most frequent type of mode is anti-symmetric stretching of the
N$_{5}^{-}$ molecule with a total of 6 modes at $\unit[990.7]{}$,
$\unit[1005.1]{}$, $\unit[1089.7]{}$, $\unit[1097.6]{}$, $\unit[1214.3]{}$,
and $\unit[1259.2]{cm^{-1}}$. There are two symmetric stretching
modes in the Raman spectra. The mode at $\unit[1200.2]{cm^{-1}}$
is the symmetric stretching of the N$_{5}^{-}$ molecule and the mode
at $\unit[2355.9]{cm^{-1}}$ is the symmetric stretching of the N$_{2}$
dimer. In comparing the Raman spectra of AgN$_{5}$-}\textit{\textcolor{black}{P2$_{1}$/c}}\textcolor{black}{{}
and AgN$_{6}$-}\textit{\textcolor{black}{P2$_{1}$2$_{1}$2}}\textcolor{black}{,
it can be seen that the crystals possess very different characteristics.
The last modes for both AgN$_{5}$-}\textit{\textcolor{black}{P2$_{1}$/c}}\textcolor{black}{{}
and AgN$_{6}$-}\textit{\textcolor{black}{P2$_{1}$2$_{1}$2}}\textcolor{black}{{}
are anti-symmetric stretching modes but differ in frequency by about
$\unit[37]{cm^{-1}}$. The most intense peak of both crystals is the
symmetric stretching mode found at $\unit[1183.4]{cm^{-1}}$ for AgN$_{5}$-}\textit{\textcolor{black}{P2$_{1}$/c}}\textcolor{black}{{}
and at $\unit[1200.2]{cm^{-1}}$ for AgN$_{6}$-}\textit{\textcolor{black}{P2$_{1}$2$_{1}$2}}\textcolor{black}{.
AgN$_{6}$-}\textit{\textcolor{black}{P2$_{1}$2$_{1}$2}}\textcolor{black}{{}
contains no bending modes compared to the six found in AgN$_{5}$-}\textit{\textcolor{black}{P2$_{1}$/c}}\textcolor{black}{.
Therefore, the correspondence between AgN$_{5}$-}\textit{\textcolor{black}{P2$_{1}$/c
crystal and that synthesized by }}\textcolor{black}{Sun }\textit{\textcolor{black}{et
al.}}\textcolor{black}{'s\cite{Sun2018} can unambiguously rule out
the appearance of AgN$_{6}$-}\textit{\textcolor{black}{P2$_{1}$2$_{1}$2
in experiment.}}

\section{\textcolor{black}{Conclusions}}

\textcolor{black}{Using first-principles evolutionary structure prediction
calculations, we discovered two novel silver pentazolate compounds:
AgN$_{5}$-}\textit{\textcolor{black}{P2$_{1}$/c}}\textcolor{black}{{}
and AgN$_{6}$-}\textit{\textcolor{black}{P2$_{1}$2$_{1}$2, }}\textcolor{black}{which
are predicted to be thermodynamically stable at pressures $\unit[41.5]{Pa}$
and $\unit[41.7]{GPa},$ respectively. These structures are proved
to be dynamically stable at $\unit[0]{GPa}$. In contrast to AgN$_{5}$
novel pentazolate AgN$_{6}$ compound contains N$_{2}$ diatomic molecules
in addition to cyclo-N$_{5}^{-}$. Synthesis of AgN$_{5}$-}\textit{\textcolor{black}{P2$_{1}$/c}}\textcolor{black}{{}
may be possible by compressing mixture of N$_{2}$ and AgN$_{3}$
to pressures above $\unit[21.42]{GPa}$ at which point AgN$_{5}$-}\textit{\textcolor{black}{P2$_{1}$/c}}\textcolor{black}{{}
is becoming thermodynamically favorable. Synthesis of AgN$_{6}$-}\textit{\textcolor{black}{P2$_{1}$2$_{1}$2}}\textcolor{black}{{}
may be possible at pressures above $\unit[25.58]{GPa}$ via the same
route. The Raman spectra of both AgN$_{5}$-}\textit{\textcolor{black}{P2$_{1}$/c}}\textcolor{black}{{}
and AgN$_{6}$-}\textit{\textcolor{black}{P2$_{1}$2$_{1}$2}}\textcolor{black}{{}
are presented. We report agreement between calculated Raman spectrum
of and experimental data from Sun }\textit{\textcolor{black}{et al.}}\textcolor{black}{\cite{Sun2018}}\textit{\textcolor{black}{{}
}}\textcolor{black}{and is therefore strong evidence that this structure
was synthesize in experiment.}
\begin{acknowledgement}
\textcolor{black}{The research is supported by DOE/NNSA (grant \#
DE-NA0003910). Computations were performed using leadership-class
HPC systems: OLCF Summit at Oak Ridge National Laboratory (ALCC and
INCITE awards MAT198) and TACC Frontera at University of Texas at
Austin (LRAC award \#DMR21006) and USF Research Computing Cluster
CIRCE.}
\end{acknowledgement}
\textcolor{black}{email: oleynik@usf.edu}

\subsubsection*{\textcolor{black}{Conflict of Interest Statement}}

\textcolor{black}{The authors declare there are no competing financial
interests.\bibliography{Silver_Pentazolate}
}
\end{document}